# Allostery without conformation change: modelling protein dynamics at multiple scales

**T C B  McLeish[1], T L Rodgers[2] and M R Wilson[1]**
[1]Biophysical Sciences Institute, Durham University, South Road, Durham DH1 3LE, UK
[2]SCEAS, The University of Manchester, Oxford Road, Manchester, M13 9PL

E-mail: t.c.b.mcleish@durham.ac.uk. Tel. +44 (0)191 334 6051

**Abstract.** The original ideas of Cooper and Dryden, that allosteric signalling can be induced between distant binding sites on proteins without any change in mean structural conformation, has proved to be a remarkably prescient insight into the rich structure of protein dynamics. It represents an alternative to the celebrated Monod-Wyman-Changeux mechanism and proposes that modulation of the amplitude of thermal fluctuations around a mean structure, rather than shifts in the structure itself, give rise to allostery in ligand binding. In a complementary approach to experiments on real proteins, here we take a theoretical route to identify the necessary structural components of this mechanism. By reviewing and extending an approach that moves from very coarse-grained to more detailed models, we show that, a fundamental requirement for a body supporting fluctuation-induced allostery is a strongly inhomogeneous elastic modulus. This requirement is reflected in many real proteins, where a good approximation of the elastic structure maps strongly coherent domains onto rigid blocks connected by more flexible interface regions.

## 1. Introduction

Allosteric signalling between distant binding-sites on proteins is a fundamental component of active biochemical networks [1]. This is the property by which the free energy of binding of a substrate at one site on a protein is modulated by the occupation or non-occupation of a distant binding site on the same protein, or within a protein complex. A successful paradigm for allostery described a structural switch on binding, through which a large-scale conformational change was triggered by the binding event, and by transmitting structural change to the second site, altered the binding affinity there in a natural way [2,3,4]. However, a large class of allosteric systems does not display any evidence of structural change during the allosteric process [5,6]. An alternative mechanism was first suggested by Cooper and Dryden [7,8], who pointed out that the thermodynamics of binding involved free energies, whose entropic component contained a contribution from thermal fluctuations of protein structure around the mean conformation. If both binding events had the effect of changing the local stiffness of the protein, and so modulating the amplitude of these fluctuations, then cooperativity between the free energies of binding was, in principle, possible. Considerable evidence has, since then, supported this mechanism for allostery in many cases, which suggests that the structural evolution of proteins has responded to this channel of information-processing in a noisy, thermal environment. One hypothesis for this coupling of allosteric sites is that binding modifies the coupling interaction between co-operative elements affecting the structural ensemble of the distant sites [9]. Here we take a different and fundamental approach, starting with the basic requirements for the realisation of allostery without conformational change, and working towards the necessary aspects of the protein structure that generate it. This is complementary to the work of analysing actual protein structures and dynamics, and provides an interpretive framework for recent and current thermodynamic and dynamic studies of real systems.

## 2. Simple Coarse-grained Models

We begin with a series of calculations that serve to illustrate and quantify the fundamental properties of fluctuation-induced allostery. These begin with extremely coarse-grained approaches, then build in progressively more detail and generality. Treating both discrete and continuum models at these levels of detail, although vastly over-simplistic in terms of real protein structure, allows us to identify how the allosteric interactions arise, and to show that they do so generally and robustly, independently of any single modelling methodology. Some have been invoked in the literature before (we point this out where this is so) but taken together as a class they are instructive in identifying the origin of the effect, the magnitudes of allosteric free energies obtainable in principle, and the structural requirements of proteins and protein complexes that generate the phenomenon. They provide an essential tool for comprehension, when faced with the much greater complexity of protein-structure based models.

*2.1. Model 1: Scissor Molecule*

We begin with the simplest possible model of a macromolecule with internal dynamics. Without loss of generality, allowing just one degree of internal dynamical freedom permits a geometric representation of a "scissor" structure as in Figure 1.

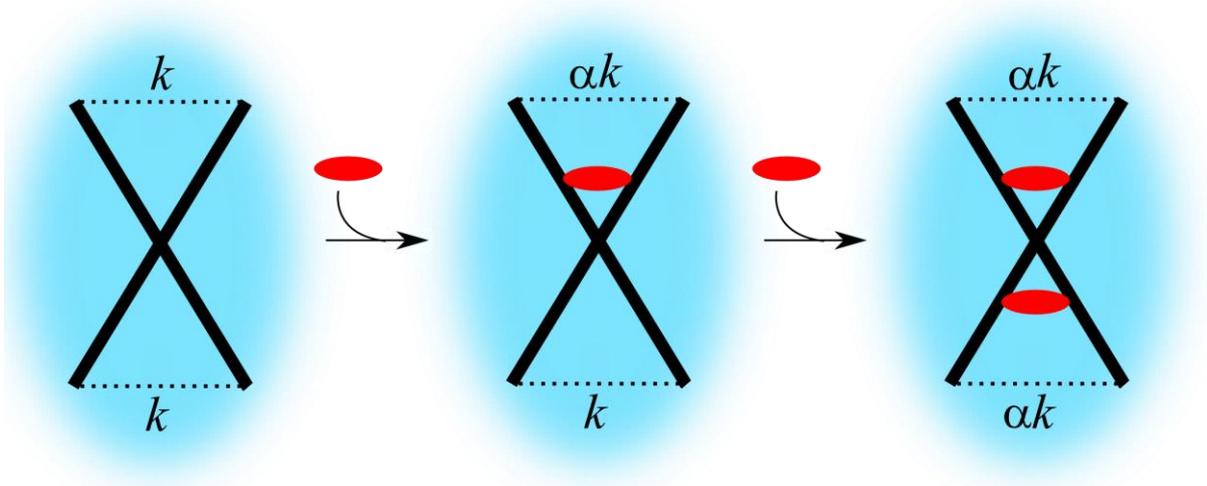

Figure 1. A scissor-molecule hinges freely around its midpoint within a harmonic potential created by stabilising interactions at its extremities. These are modulated locally by binding to substrates, changing their effective spring constants, $k$, by a factor $\alpha$.

The (classical) free energy of a simple degree of freedom in a harmonic potential is straightforward to calculate in terms of the partition function $G = -k_B T \ln Z$ where the partition function is integrated over the spatial coordinate of the potential of effective spring constant $k$:

$$Z = \frac{1}{h}\int_{-\infty}^{\infty} e^{-kx^2/2}\,\mathrm{d}x = \frac{1}{h}\sqrt{\frac{2\pi}{k}} \tag{1}$$

The free energy depends logarithmically on the strength of the elastic potential for the global "scissor-mode" for this model molecule. The curvature of the log function is the key to the possibility of allosteric signalling on binding (Figure 2a). We suppose that a binding site at each extremity of the scissor molecule supports a substrate, and that this in turn changes the *local* contribution to the spring constant for the mode from $k$ to $\alpha k$. The free energy increases on binding with $\ln(k_{TOT})$ as the entropy of thermal fluctuations decreases. But as the total spring constant increases, so the contribution to the free energy from equal increases in $k_{TOT}$ reduces. In particular the entropic free energy penalty of binding the second substrate is less that than of the first, implying a natural positive cooperativity of binding. It is straightforward to calculate the allosteric free energy, $\Delta\Delta G$:

$$\Delta\Delta G = \frac{1}{2}k_B T \ln\frac{2k(2\alpha k)}{(k+\alpha k)^2} = \frac{1}{2}k_B T \ln\frac{4\alpha}{1+2\alpha+\alpha^2} < 0 \tag{2}$$

Where, if $\Delta G_1$ and $\Delta G_2$ are the binding free energies of the first and second substrates, $\Delta\Delta G=\Delta G_2 - \Delta G_1$. Note that the absolute stiffness of the global mode does not appear in the expression for the allosteric free energy, which depends only on the dimensionless change in local stiffness, $\alpha$.

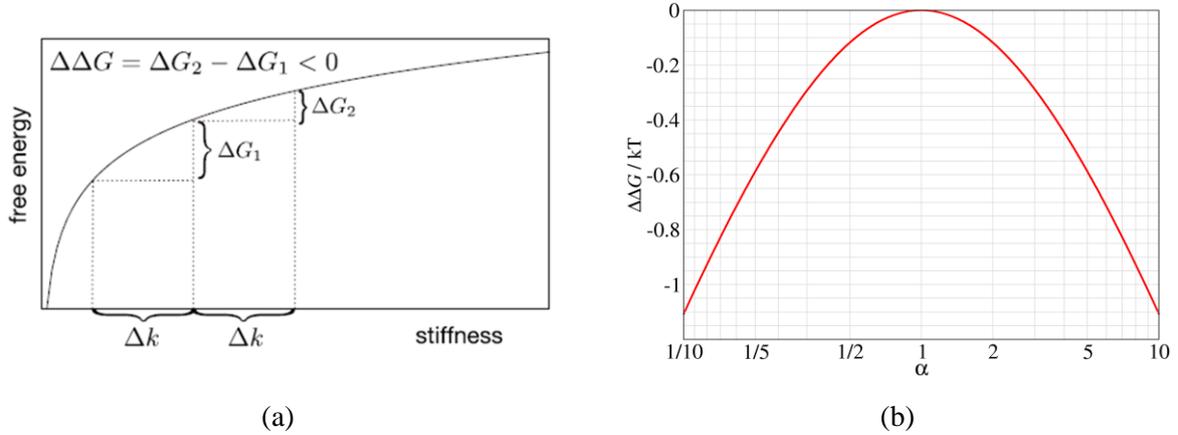

(a) (b)

Figure 2. (a) The entropic part of the free energy of a single harmonic mode of thermal fluctuation in terms of the effective stiffness, $k$, of the mode. Increasing stiffness by equal sequential increments results in a smaller increment on the second augmentation of stiffness; (b) the allosteric free energy as a function of the dimensionless local stiffness change on each binding site, $\alpha$.

This result is plotted in Figure 2(b). The allosteric free energy is negative (co-operative) whether the binding stiffens or weakens the local stiffness contribution to the mode, and diverges logarithmically (so weakly) as the stiffness change increases. For reasonable (though non-perturbative) values of $\alpha$ (factors no larger than 3) that would be consistent with, for example, the establishment of one or two additional hydrogen bonds at the location of a binding site, we observe that the contribution of just a single mode to the allosteric free energy is of the order of $0.3 k_B T$. This means that with only a few modes active in this manner across a protein, signalling free energies of a few kcal/mol can be achieved. The benefit of this very simple "toy" model is not to model any particular protein in particular, even in a maximally coarse-grained sense, but to show that, in principle, moderate modifications of the thermal fluctuations in structures, that generate coherent dynamics globally across a protein *via* normal modes, are candidates for significant allosteric free energies without conformational change. We also note that the effect, while depending crucially on the thermal fluctuating dynamics of the system, is an equilibrium effect, and is independent of type of the dynamics (overdamped, or underdamped) or on details of correlation functions.

*2.2. Model 2: A modified Scissor molecule with flexible hinge*

In general, a large molecule has many degrees of freedom that may be coupled in complex ways. However, if the couplings are all harmonic (quadratic in the coordinates) then the Gaussian partition function and free energy of the single mode generalises in a straightforward manner. In the general case, the full Hamiltonian is

$$\mathsf{H} = \frac{1}{2}\mathbf{p}^T \mathbb{M}^{-1}\mathbf{p} + \mathbf{x}\mathbb{K}\mathbf{x} \qquad (3)$$

where **p** is the vector of the momenta and **x** that of coordinates. $\mathbb{M}$ and $\mathbb{K}$ are the (diagonal) mass matrix and (symmetric) spring coupling matrix respectively.

Models of just this type are generated as realistic representations of proteins through the Elastic Network Model (ENM) approach [10,11]. The entropic free energy becomes a matrix determinant:

$$G = -k_B T \ln Z = \frac{1}{2} k_B T \ln |\mathbb{K}| + \text{const} \qquad (4)$$

and the allosteric free energy, a logarithmic function of the determinants of the many-body harmonic system with zero, one and two effectors bound.

$$\Delta\Delta G = \frac{1}{2} k_B T \ln \frac{|\mathbb{K}_0||\mathbb{K}_2|}{|\mathbb{K}_1|^2} \qquad (5)$$

New features of a dynamically allosteric system are generated by increasing the number of degrees of freedom above one; these new features become apparent even when the number of degrees of freedom is not increased by a large amount. Even the modified, and rather more realistic, "scissor" model in which the hinge is replaced by a third spring rather than a stiff pivot, displays some surprising features, Figure 3, increasing the number of degrees of freedom to two. This model was discussed previously in an approach to proteins (such as the *lac* repressor) in which relative position and orientation of two inherently-stiff subdomains dominates the low-frequency motion [12].

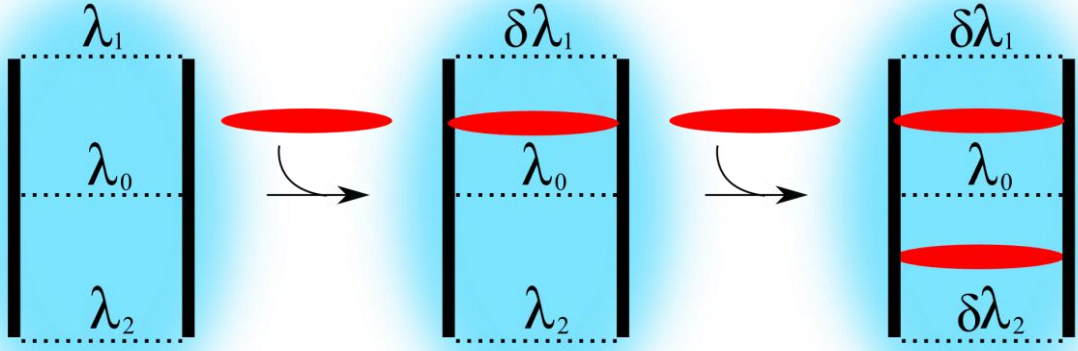

Figure 3. A modified scissor-molecule with central hinge replaced with a flexible spring. These are modulated locally by binding to substrates, changing their effective spring constants, $\lambda$, by a factor $\delta$.

In addition to the displacement coordinate between the two rigid rods, $x$, there is also now an angular displacement between the two rods away from their parallel equilibrium, $\theta$. The vector of generalised coordinates, the potential energy matrix and the allosteric free energy of the system has the structure of eqn (6):

$$\mathbf{x} = \begin{pmatrix} x \\ \theta \end{pmatrix}$$

$$\mathbb{K} = \begin{pmatrix} \lambda_1 + \lambda_2 + \lambda_0 & \lambda_1 - \lambda_2 \\ \lambda_1 - \lambda_2 & \lambda_1 + \lambda_2 \end{pmatrix} \quad (6)$$

$$\Delta\Delta G = \frac{1}{2} k_B T \ln\left( \frac{(4\delta + 2\lambda_0\delta)(4\delta + \lambda_0)}{[4\delta + \lambda_0(1+\delta)]^2} \right)$$

$$= 0 \quad \text{when } \lambda_0 = 0$$

Here the spring constants at the binding sites $\lambda_1$ and $\lambda_2$, are set (without loss of generality) to 1 and are enhanced by a dimensionless increment $\delta$ on substrate binding. The dimensionless hinge spring constant $\lambda_0$ becomes a second parameter in the now two-dimensional landscape for $\Delta\Delta G$. Remarkably, however, if a large value of the substrate effect $\delta$ is chosen and the central hinge stiffness is set to zero, the allosteric effect vanishes. As the central hinge stiffness increases, the result of model 1 is asymptotically approached. Other cases of a null effect also appear, counter-intuitively, as we see in the next case.

Perhaps the most interesting feature of the "flexible hinge" model is, however, the appearance of a regime of negative cooperativity within the parameter space of the model. The full landscape of $\Delta\Delta G(\lambda_0, \delta)$ is plotted in Figure 4. Unlike the limiting case (at high $\lambda_0$) of uniformly negative $\Delta\Delta G$ (and so positive binding cooperativity), here a region of positive $\Delta\Delta G$ (negative cooperativity) appears when the effect of binding is to stiffen. This is a very powerful property of the fluctuation mechanism – it is able to account for either sign of cooperativity in a natural way.

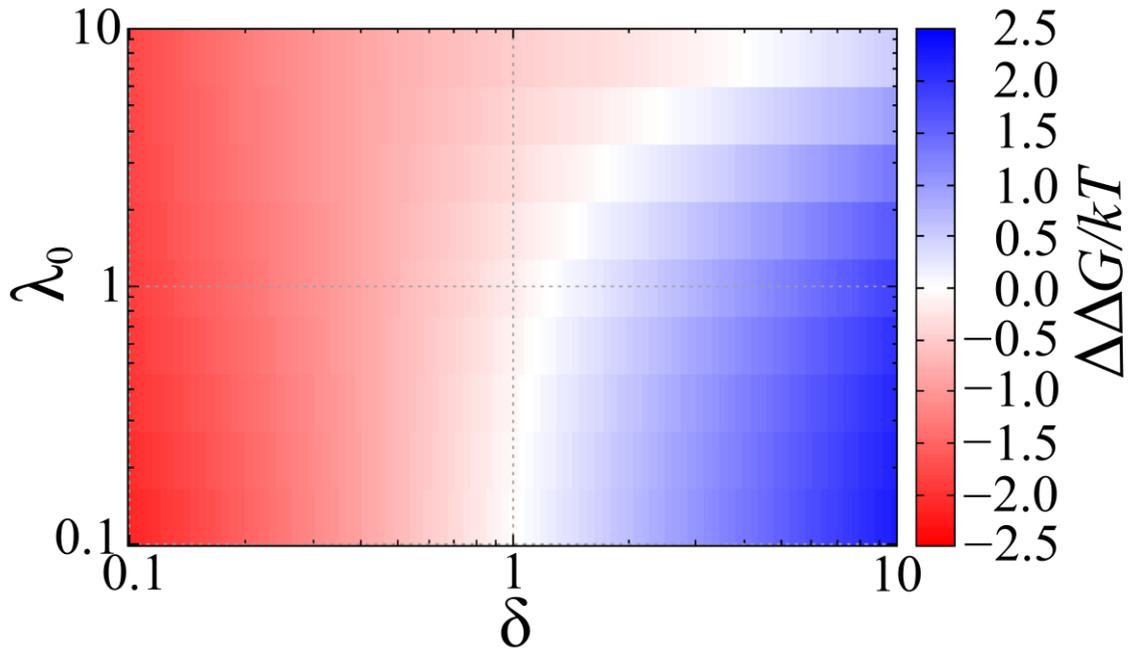

Figure 4. The allosteric free-energy landscape of the flexible hinge model. The allosteric free energy $\Delta\Delta G$ is indicated by colour as a function of the two dimensionless model parameters of the central spring constant $\lambda_0$, and the local stiffness modification on binding $\delta$.

The underlying physics of negative fluctuation-induced allostery can be understood from the structure of the normal modes. When there is just a single degree of freedom, or when all are uncoupled, then changing an effective stiffness does not alter the eigenvectors of the elastic modes – the structure of the modes is unchanged. There is in this case no alternative to the positive cooperativity induced by the curvature of the log function of figure 2. However, once there are two or more coupled degrees of freedom, then the local alteration of stiffness can have an additional effect beyond that of simply stiffening a normal mode – the eigenvector structure of the normal modes can itself be altered. This in general reduces the entropic penalty on the first binding from the case of fixed mode structure. Alternatively, amplitude-strength is shifted towards modes less affected by the local change of stiffness.

*2.3. Model 3 Continuum Elasticity*

If discrete elastic systems show a fluctuation-induced allosteric signal when the elastic potential is modified locally (e.g. at a hinge in the scissor model), it is informative to examine the continuous case as well. Does the property extend to continuous elastic media? It seems likely *a priori*, that it might, given that, just as in the discrete case, elastic continua support normal modes of motion at many different characteristic wavelengths. Of course, a normal mode analysis of a continuum body maps it into a discrete dynamical system (of a degree depending on the high-frequency cut-off), but examining the continuum case provides both greater robustness to our understanding of the mechanism, and also further insights into the conditions under which it arises.

By modifying the elastic constants locally, long-wavelength modes (those with finite amplitude at the modified site) will themselves be perturbed. A second modification at another location where such a mode has amplitude might be expected to deliver a non-zero allosteric free energy, for the reason identified through the simple models studied so far. While appearing a rather abstract question, it is relevant to proteins since the fluctuation-induced allosteric effect relies on long-wavelength, global modes of motion across the protein. These very motions reflect an elastic structure that is itself coarse-grained to the relevant lengthscales. Molecular detail might be expected to be averaged-over sufficiently that the model of an elastic continuum becomes valid. Furthermore, it has been known for some time that the mode structure of protein dynamics is reproduced faithfully at low and intermediate frequencies by elastic network models that use an invariant spring constant between nearby Cα carbons [10,12].

The appropriate calculation in this case is in second order perturbation theory [13]. An elastic partition function is calculated in a normal mode expansion of the form of equations (4) and (5) above, where $\mathbf{K}_1$ and $\mathbf{K}_2$ are the elastic energy matrices of the perturbed system with enhanced elastic constants local to two distant positions in the continuum $r_1$ and $r_2$. To make this concrete we take the 1-dimensional case of scalar elasticity, with a Hamiltonian of

$$\mathsf{H} = \frac{1}{2}\int \kappa(x)(\nabla_x \phi)^2 \, dx \tag{7}$$

where $\phi(x)$ is a continuous displacement field. The elastic modulus $\kappa(x)$ is constant apart from the local perturbations, $\tilde{\kappa}$, on binding consecutive substrates ($\kappa_1$ then $\kappa_2$) at the binding sites:

$$\begin{aligned}\kappa_1(x) &= \kappa_0 + \tilde{\kappa}\delta(x-x_1) \\ \kappa_2(x) &= \kappa_0 + \tilde{\kappa}[\delta(x-x_1) + \delta(x-x_2)]\end{aligned} \tag{8}$$

The evaluation of the free energy (which in this case should strictly be the Helmholtz free energy $F$) to second order in the perturbation is straightforward as an expansion of equation (4):

$$F = F_0 - k_B T \left(1 - \beta \langle \mathsf{H}_i \rangle_0 + \frac{1}{2}\beta^2 \left( \langle \mathsf{H}_i^2 \rangle_0 - \langle \mathsf{H}_i \rangle_0^2 \right)\right) \tag{9}$$

(using the usual notation of $\beta=1/k_B T$). The allosteric free energy is likewise found as an expansion to second order of equation (5) for the case of the perturbation of equation (8) directly in terms of a correlation function under the unperturbed case:

$$\Delta\Delta F = -\frac{\beta}{4}\tilde{\kappa}^2 \left[ \left\langle (\nabla_x \phi(x_1))^2 (\nabla_x \phi(x_2))^2 \right\rangle_0 - \left\langle (\nabla_x \phi(x_1))^2 \right\rangle_0 \left\langle (\nabla_x \phi(x_2))^2 \right\rangle_0 \right] \tag{10}$$

This result is general and independent of the form of the Hamiltonian in equation (7), providing that the perturbation has the local structure of equation (8). It is simple to evaluate equation (10) for the specific, and simplest, case of equation (7), which describes a continuum with local stretching energy only. The normal modes are simply the harmonic functions of a Fourier expansion of $\phi(x)$, $\phi_q(x)=e^{iqx}$. The amplitude of each mode's contribution to the final expression, equation (10), is governed by the balance of the quadratic dependence on the gradient of the displacement (the local stretch) in equation (7) and the square-gradient terms in the correlation function, equation (10). The result is that each harmonic mode is equally weighted, and so the spatial dependence of the allosteric response is entirely local:

$$\Delta\Delta F(x_1, x_2) = -\frac{\tilde{\kappa}^2}{\kappa_0}\delta(x_1 - x_2) \tag{11}$$

The surprising result of this analysis is that there is no long-range fluctuation allostery at all. The structure is no different in higher dimensions: with simple local elastic stretch penalised, the effect is entirely local. This is a consequence of the absence of any characteristic lengthscale within the elastic energy in this simple case. A long-range interaction requires such a lengthscale, which itself may arise in two ways. Firstly, the background elastic modulus might not be constant, but carry local variations. This is in fact how the simple models in the first two cases above work. The scissor-structure underlying them is, in this light, an extreme case of elastic inhomogeneity: the rod-like domains linked at the pivot point are of course themselves elastic bodies, but with their internal elastic modulus taken infinitely larger than that arising from their connecting interface. In this way the system circumvents the initially restrictive result of equation (11). A second route to finite-range fluctuation allostery is by invoking a more structured elastic Hamiltonian with inherent length scales. A natural route here is to include bending, as well as stretching, elasticity. In our example, such a modification takes the form:

$$H = \frac{1}{2}\int \left[\kappa(x)(\nabla_x \phi)^2 + \gamma(x)(\nabla_x^2 \phi)^2\right]dx \qquad (12)$$

where $\gamma$ is the bending elastic constant. In this case, carrying out the calculation in second order perturbation as before, keeping the same perturbation, i.e. only perturbing the stretching modulus on binding, not the local bending modulus, yields:

$$\Delta\Delta F = -\tilde{\kappa}^2 \left|\int dq\, q^2 \frac{1}{\gamma q^4 + \kappa_0 q^2} e^{iq(x_1-x_2)}\right|^2$$

$$= -\frac{\tilde{\kappa}^2}{\gamma \kappa_0}\left[e^{-|x_1-x_2|/\xi}\right]^2 \qquad (13)$$

Now we do have a finite-range allosteric effect, with the length scale of the elastic "screening length" $\xi = \sqrt{(\gamma/\kappa_0)}$. This result holds in all dimensions. An example of this effect was identified in the example of one-dimensional systems in a model for the allosteric binding of the motor protein dynein to microtubules [14] whereas the bending flexibility of the twisted helical coiled-coil of the binding domain increased, the allosteric free energy of binding to the microtubule decreased.

*2.4. Model 4: Local dynamical coupling and entropy-enthalpy compensation*

The thermal fluctuations of protein structure are, of course, not confined to the slower, longer range, modes of motion, although these are naturally the ones involved in generating induced allostery by virtue of their long-range correlations. More local, higher-frequency motions of both main-chain conformation and side-chain motion are evidenced directly in NMR NOE assays [5] and in molecular-dynamics simulation [15]. It might be thought that such local and high-frequency motion could play no part in allosteric signalling, but in many cases the thermodynamics suggests otherwise. The very simple, coarse-grained models we have examined so far suggest that the allosteric free energy of systems without conformational change, but which rely purely on the modulation of long-range-correlated thermal fluctuations, have a signature of largely entropic contributions. The enthalpic binding energies required to generate substrate binding in the first place would contribute equally at both sites, and so would not appear in the allosteric (difference) free energy. Furthermore, with no more than an order of 10 long-range modes (including the "polarisation" of the tensorial elastic modes in three dimensional structures) available, the entropic free energies would be of the order of a few $k_BT$. There are indeed cases in which the entropic term dominates the allostery [16], but others in which although the net $\Delta\Delta G$ is co-operative and of this order of magnitude, it contains largely self-compensating entropic and enthalpic terms of the order of 20 or 30 $k_BT$ [6]. Note that these are cases where no structural change is evidenced, so that an entropic effect from the exposure of water to greater or lesser exposure of hydrophobic surfaces is not a candidate for the entropic changes on substrate binding. Somehow, faster, more local fluctuations are appearing within the allosteric part of the free energy, but are attracting compensating enthalpies in these cases.

A simple but plausible model which possesses just this property invokes a coupling between the amplitude of global and local (e.g. side-chain) dynamics [17,18]. If large-scale fluctuations are present at high amplitude, this is tantamount to saying that a high degree of disorder is present in the protein. In such a case, local energy-favourable structural contacts between side-chains will be compromised: the mean native structure will be visited for only a fraction of a time-series of structures. As a consequence, the time-averaged enthalpy of side-chain interactions will be higher, and the amplitude of their structural fluctuations will also be high. They will also be more exposed to the solvent. This effect can of course contribute with either sign, but in the case of hydrophobic side-chains will also contribute to a higher enthalpy when the global modes are active. There is a correlation between the entropy of global and local modes of motion when such a coupling is present. However, when fluctuations around the mean native structure are small, the environment of side chains will be more stable, and their local interactions will be satisfied for more of the time. In consequence the side chain motion will also be more restricted, and their contribution to the total entropy will also be lower.

Just as in the simple models 1 and 2, it is important to analyse physical insights like these quantitatively. A simple approach is illustrated in Figure 5.

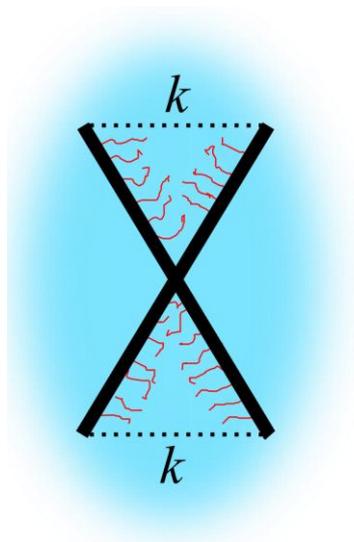

Figure 5. A version of the scissor model with coupling to local (side chain) modes. Small scale structures "ride" on the larger structures. When the amplitude of the global modes is high, the environment of the local structures is more unstable and disordered, inducing higher entropy and less negative enthalpic contributions to the total free energy from their structure and dynamics.

The mathematical structure of a harmonic potential for the fast, local modes that captures this property takes the following form:

$$V_{f_i}(x_{f_i}, x_s) = -V_{f_0} f_i(x_s) + \frac{1}{2} g_i(x_s) k_{f_i} x_{f_i}^2 \tag{14}$$

The two functions $f_i(x_s)$ and $g_i(x_s)$ describe the dependence of the depth and strength of the fast-mode localising potentials on the amplitude of the (single in this case) slow mode $x_s$. The qualitative features of a mode-coupling model such as this one will not depend on the exact form of these coupling functions, so it is sensible to choose functional forms that allow direct calculation of the thermodynamic structure. A suitable model form is

$$V_{f_i} = -V_{f_0} f_i \left( -\frac{k_v x_s^2}{kT} + 1 \right) + \frac{1}{2} \left( \frac{k_f}{\exp\left(\frac{k_k x_s^2}{2kT}\right)} \right) x_{f_i}^2 \tag{15}$$

Note that we assume that the couplings, parameterised through $k_v$ and $k_k$ (which are combined with the thermal energy in equation (15) to give them the same dimensions as the spring constants for slow and fast modes $k_s$ and $k_f$), are identical for the $N$ fast modes. $N$ becomes the other new parameter for the coupling model. Working through the calculation of the allosteric free energy in the same manner as for Model 1 gives a new expression:

$$\Delta\Delta G = \frac{k_B T}{2} \ln\left( \frac{(\alpha^2 + NA)(1 + NA)}{(\alpha + NA)^2} \right) \quad ; \quad A = k_v V_0 - k_k \tag{16}$$

Here $\alpha$ has the same meaning as in Model 1. Intriguingly the structure of the allosteric free energy is little different from that of the simpler model with the bare global mode active. The number of fast modes $N$ does enter, but only logarithmically at the level of $\Delta\Delta G$.

However, this is not true of the entropic and enthalpic contributions to $\Delta\Delta G$. Making an explicit evaluation of the allosteric enthalpy:

$$\Delta\Delta H = k_B T N k_k \ln\left(\frac{1}{\alpha^2 + NA} - \frac{1}{\alpha + NA} + \frac{1}{1 + NA}\right) \quad (17)$$

reveals a term that scales directly with $N$. Of course, given the absence of such a term in the expression for $\Delta\Delta G$, is it clear by construction, $\Delta\Delta G = \Delta\Delta H - T\Delta\Delta S$, that the entropy must indeed possess a compensating term of equal magnitude. When this model is parameterised to real dynamically allosteric proteins, $N$ typically takes a value of order 10 [19]. It remains a challenge to identify this effect explicitly in fully-atomistic simulations or in NMR data, although in the experimental case it is observed that low frequency and high frequency motions are correlated in amplitude when these differ between the apo and holo state of a protein [5].

### 3. An Elastic Network Model of Fluctuation-induced Allostery

The value of the artificially simple models above is in their ability to capture the origin and qualitative structure of the dynamic allosteric effect, not in their description of any real system. To do that, finer-grained calculations are required that themselves respect real protein structure at an appropriate level of detail. They will also validate or invalidate the hypotheses of the simple models as finer detail is introduced. There are at least two further levels of detail that can be introduced, at the penalty at each stage of heavier calculation and consequently harder limits on the time intervals accessible to computation.

A powerful compromise method, capturing the elastic structure at the level of protein domains but not requiring the computation of side-chain motion at atomistic detail, is the elastic network model (ENM). Here, every Cα carbon is connected to all others within a cut-off radius, $R_c$, (usually of the order of 10Å) with a harmonic spring of uniform stiffness, $k_{ij}$,

$$V_{ij} = \begin{cases} \dfrac{k_{ij}}{2}(r_{ij} - R_{ij})^2 & R_{ij}^2 \leq R_c^2 \\ 0 & R_{ij}^2 > R_c^2 \end{cases} \quad (18)$$

where $R_{ij}$ is the equilibrium distance between the atoms, $r_{ij}$ is the distance between the same atoms, and $V_{ij}$ is the potential energy of the spring. This then can be used to produce the force matrix for the system, the mass weighted Hessian matrix, $\mathbf{D}_{ij}$,

$$\mathbf{D}_{ij} = \frac{1}{2\|\mathbf{x}_j - \mathbf{x}_i\|^2}(\mathbf{x}_j - \mathbf{x}_i)^T m_i^{-1/2} k_{ij} m_j^{-1/2} (\mathbf{x}_j - \mathbf{x}_i) \quad (19)$$

where $\mathbf{x}$ is the atom coordinate and $m$ is the atom mass. The Gaussian system generated can then be analysed via its normal mode structure (diaganalising the Hessian matrix), and entropy changes calculated on substrate binding. At the level of the ENM, the binding event is modelled by a local increase in the density of harmonic springs at the binding site (one new site per ligand at the average coordinate). We have carried out ENM computations (ran with ΔΔPT [20]) of this nature on the Catabolite Activator Protein (CAP) of Escherichia coli that binds cAMP generated by adenylyl cyclase in response to the phosphorylated form of Enzyme IIAGlc (phosphorylated in response to the phosphoenolpyruvate-carbohydrate phosphotransferase system) [21,22,23].

CAP is known to show mild dynamically-driven negative allostery between the two identical binding sites for cAMP. The ENM predictions for the unbinding constants also shows a mild negative cooperativity (from equation 20 where $v$ is the eigenvalues of $\mathbf{D}_{ij}$ and 0, 1, and 2 refer to the *apo*, *holo1*, and *holo2* states of the protein respectively) in qualitative agreement with experiment (ratios of unbinding constants are 1.6 in isothermal calorimetry experiment and 1.35 in the ENM (using the first 100 modes)).

$$\text{coop} = \prod_{i=1}^{100} \frac{v_{i,0} v_{i,2}}{v_{i,1}^2} \tag{20}$$

Experimental and theoretical calculations at very coarse-grained levels have shown that this effect arises from a slight decrease in fluctuation on the first binding event (sometimes even an increase), followed by a larger decrease on the second. It therefore falls into the class of entropically-generated negative allostery captured qualitatively by the simple flexible-hinge model of section 2.2, though is in detail more complex.

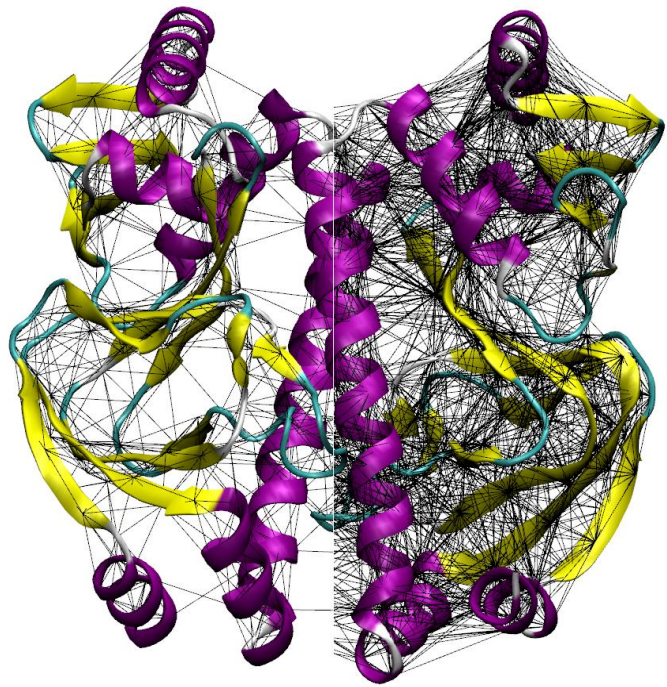

Figure 6. ENM models for the CAP homodimer using 8Å (left half) and 12Å (right half) cut-offs for the insertion of Cα-coupling harmonic springs. The extremely inhomogeneous elastic density is clear in either representation.

By summing over normal modes the local motion amplitude of each residue can be calculated. This quantity bears experimental comparison to the "B-factor" values from X-ray crystal data. The comparison of B-factor data from the ENM is remarkably accurate at either of the resolutions of seen in Figure 6 (see Figure 7 for the apo case).

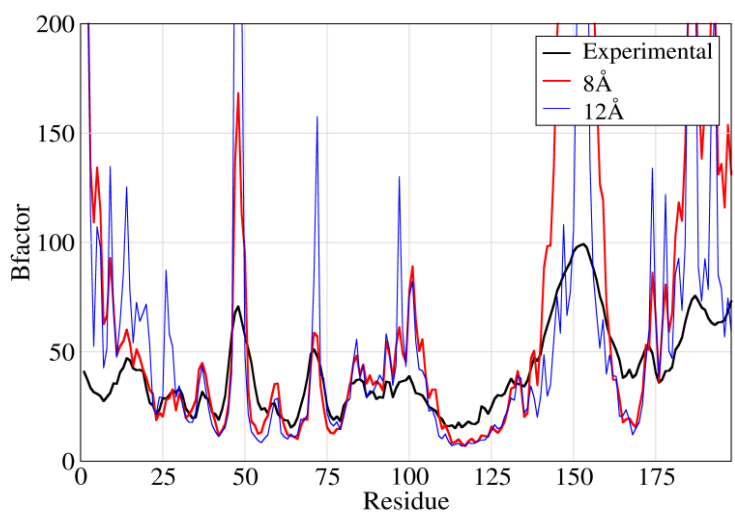

Figure 7. B-factor predictions from both cut-off choices of ENM of Figure 6, compared with experimental data from X-ray structure determination of CAP.

The crystal environment has a natural effect of constraining the amplitude of the most freely-moving residues, but the detailed structure of the internal protein dynamics is captured. Furthermore the successive binding of cAMP to the binding sites generates a change in dynamics (accessed through predicted B-factors) which is consistent with the explanation the negative co-operative thermodynamics in terms of fluctuation-amplitude entropy changes – there is a strong correlation across point-mutations between the degree to which they modulate fluctuations at the cAMP binding site and their thermodynamic cooperativity.

The resilience of the ENM level of description can be supported by finer-grained molecular dynamics as well as by experiment. Atomistic MD is able to validate ENM predictions for local residue motions outside of the experimental crystal environment.

## 4. Rotational Translational Block (RTB) Description

The highly inhomogenous density of internal Cα-connecting springs generated by the ENM suggests that an approximation to the dynamical structure identifies the very dense regions, and approximates them as rigid blocks, with their internal residues fixed in relation to each other. Each block is then connected to its neighbours by the remaining springs in the less-dense regions between them. It is possible to do this in a regularised way [24]. Each rigid block is able to translate and rotate relatively to the others in the harmonic potentials generated by the interfacial springs. The number of degrees of freedom of this system is vastly smaller than that of even the full ENM, so very efficient to calculate with:

$$\mathbf{D}^{rtb} = \mathbf{P}^T \mathbf{D} \mathbf{P}$$

$$\mathbf{P}^{\mu}_{j,i\nu} = \begin{cases} \sqrt{\dfrac{m_i}{M_j}} \delta_{\mu\nu} & \mu = 1,2,3 \\ \sum_{\alpha\beta} \sqrt{\dfrac{m_i}{I_{j,\mu-3,\alpha}}} (r_i - r_j^0) \varepsilon_\alpha \beta_\nu & \mu = 4,5,6 \end{cases} \quad (21)$$

where the RTB-Hessian, $\mathbf{D}^{rtb}$, is of size 6nb × 6nb instead of 3N × 3N matrix ($\mathbf{D}$ given from equation 19), where the reduction vector, $\mathbf{P}$, has indexes μ (the translation (1, 2, and 3) and rotation (4, 5, and 6) of each block), $M_j$ is the block mass, $I_j$ is the block moment of inertia, $r_j$ is the centre of mass of the block, $\delta_{\mu\nu}$ is biorthogonal vector between the two spaces and ε is the Levi-Civita symbol.

We have applied the RTB analysis to the CAP dimer, with the result that an accurate description is captured with four rigid domains (two for each monomer – see Figure 8)

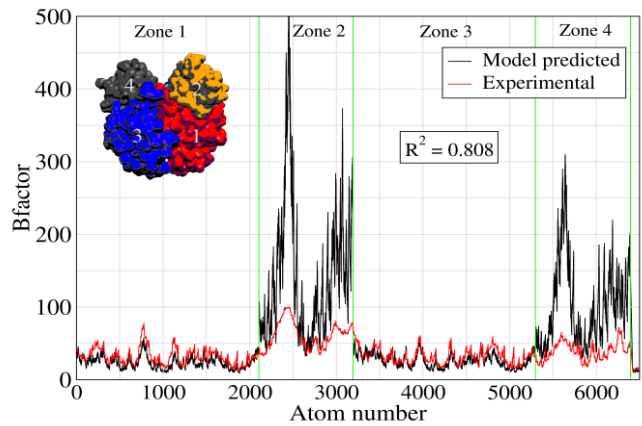

Figure 8: The RTB model for the CAP homodimer. The four rigid domains are indicated in the insert, and the resulting B-factor prediction compared with experiment in the same way as for the ENM in Figure 7.

Remarkably, the B-factor distribution is almost as accurately predicted by the RTB model as by the full ENM. This is true even of detailed differences in fluctuation amplitude within the rigid blocks. This surprising observation implies that the dominant reason for higher or lower amplitude of RMS fluctuation in the spatial position of a residue is not its local environment, but its distance from the centre of rotation of the effective block in which it is embedded. The cooperativity predicted by this model is also in agreement with the experimental value by predicting slight negative cooperativity (1.08).

## 5. Conclusion

Fluctuation-induced allostery in proteins requires interactions over a wide range of length scales to be understood. Highly coarse-grained models provide insight into how the phenomenon arises, including the requirement for inhomogenous elastic density. This arises both through the co-existence of relatively rigid and less rigid regions within the protein, and through the complex boundary geometry of the folded protein (one can think of voids within the overall region of the protein as regions of vanishingly low elastic modulus). This underlying physical constraint suggests explorations of the evolutionary pressures on inhomogenous structures in proteins.

One significant advantage of fluctuation-induced allostery is that it can generate cooperativity of binding at long range of either sign. As long as the dynamical system at long range contains at least two coupled degrees of freedom, there are regions in parameter space which produce both positive and negative allostery. This occurs through the modification of the normal mode structure on binding.

The structure of modes, and the nature of their modification is important in any case to support the coarse-grained models we have suggested capture the mechanism correctly. One objection to the coarse-grained approach might be that they do not account for extra degrees of freedom when a substrate molecule is bound [25]. Of course, in the ENM models the network is enhanced by the addition of an extra node on modelling the substrate binding, so that the effect is taken into account. But this does not happen in the very coarse-grained models. The justification here is that the new degrees freedom which appear are always *local*, and therefore unable to transmit the long-range allosteric effect. It is only the modification to the structure and stiffness of the pre-existing long range modes that conveys the allosteric effect, and this is captured by the coarse-grained models.

Coupling between large-scale global modes and local, high-frequency, modes gives rise to compensating entropic and enthalpic terms which can be much larger than the overall allosteric free energy (since only the compensating contributions scale as the number of coupled local modes). Finer scaled models allow representation of detailed protein structure as specified by NMR or X-ray analysis. The ENM approach gives a faithful account of binding thermodynamics and the amplitude of local residue fluctuation. It also suggests a coarse-grained model tailored to each individual protein in which more elastically-dense regions are replaced by rigid blocks, free only to rotate and translate relative to each other. Even a few blocks can give a remarkably accurate representation of protein dynamics.

There is a further, highly suggestive, consequence of this observation, which makes contact between this finding, generated by fine-grained approaches to protein dynamics, and the very coarse-grained simplistic models with which we began. The coarse-grained theories led us to the conclusion that, in order to generate the capacity to signal allosterically between distant binding sites, a protein requires a highly anisotropic internal elastic structure. The RTB analysis of real allosteric proteins shows that, qualitatively, this is precisely the internal engineering that their evolution has driven them to adopt. As a result, the apparently wildly over-simplistic block-elastic models are in actuality not very wide of the mark as realistic coarse-grained pictures of real proteins.

**Acknowledgements**

We thank M. Cann, D. Burnell, P. Townsend, E. Pohl, A. Levine and R. Bruinsma for helpful discussions during this work, and P. Townsend and E. Pohl for the B-factor data on the CAP homodimer. This work was supported by EPSRC grant EP/H051759/1.